\documentclass[12pt]{article}

\begin{document}

\begin{titlepage}

%\begin{flushright}
%hep-th/0402??? \\
%\end{flushright}

\vskip 1cm
\begin{center}
{\large\bf 4D equivalence theorem and gauge symmetry on orbifold}

\vskip 1cm {\normalsize Y.\ Abe,
N.\ Haba*, K.\ Hayakawa, Y.\ Matsumoto, M.\ Matsunaga and K.\ Miyachi}
\\
\vskip 0.5cm{\it Department of Physics Engineering, Mie University, Tsu, Mie,
    514-8507, Japan}
\\
\vskip 0.5cm *{\it Institute of Theoretical Physics,
  University of Tokushima, Tokushima 770-8502, Japan}
\end{center}
\vskip .5cm
\vspace{.3cm}

%%%%%%%%%%%%%%%%%%%%%%%%%%%%%%%%%%%%%%%%%%%%%%
%%%%%%%%%%%%%%  ABSTRACT %%%%%%%%%%%%%%%%%%%%%
%%%%%%%%%%%%%%%%%%%%%%%%%%%%%%%%%%%%%%%%%%%%%%
\begin{abstract}

We investigate the high-energy behavior  of
    the scattering amplitudes in
    the extra dimensional gauge theory
    where the gauge symmetry is broken
    by the boundary condition.
We study, in particular,  the 5-dimensional
    $SU(5)$ grand unified theory
    whose 5th-dimensional coordinate
    is compactified on $S^1/Z_2$.
We pay attention to
    the  gauge symmetry
    compatible with the boundary
    condition on the orbifold
    and give the BRST formalism of the 4D theory
    which is obtained through integration of the 5D theory along the
    extra dimension.
We derive the 4D equivalence theorem
    on the basis of the Slavnov-Taylor identities.
We  calculate the amplitudes of the process
    including four massive gauge bosons
    in the external lines and compare them with the ones
    for the connected reactions  where the gauge fields are replaced
    by the corresponding would-be NG-like fields.
We explicitly confirm the equivalence theorem to hold.

\end{abstract}
\end{titlepage}

%%%%%%%%%%%%%%%%%%%%%%%%%%%%%%%%%%%%%%%%%%%%%%%%%%%%%%%%%%%%%%%%%%%
%%%%%%%%%%%%%%%%%%%%% INTRODUCTION %%%%%%%%%%%%%%%%%%%%%%%%%%%%%%%%
%%%%%%%%%%%%%%%%%%%%%%%%%%%%%%%%%%%%%%%%%%%%%%%%%%%%%%%%%%%%%%%%%%%
\section{Introduction}

It is well-known that
  in the 4D gauge theories
  with the explicit gauge
  symmetry breakings,
  the amplitudes of four massive gauge bosons
  in the external lines
  show the bad high-energy behavior
  $O(E^4/m^4)$ and $O(E^2/m^2)$,
  which
  breaks the unitarity.
On the other hand,
  when the gauge bosons obtain
  their masses through
  the Higgs mechanism,
  the power law behavior, $O(E^4/m^4)$ and $O(E^2/m^2)$,
  is canceled
  by the contribution of the Higgs bosons
  \cite{LlewellynSmith, Dicus,
  GBequivalence, SMunitarity, Chanowitz}.
The cancellation is guaranteed to occur
  by the equivalence theorem
  which states that
  the amplitude of massive gauge bosons
  in the external lines is the same, up to some constant facotor, as
  that of  the connected reaction
  where the gauge fields are replaced
  by the corresponding would-be NG-like fields
  \cite{Gounaris, Yao, Bagger, He}.

Then, what is going on
  the unitarity
  in the extra
  dimensional gauge theories
  where the gauge symmetries
  are broken by the boundary
  conditions?
In the higher dimensional gauge theory,
  the reduction
  of gauge symmetry
  is realized
  through the boundary condition
  of the extra dimensional
  coordinate (see, for examples, \cite{5d}).
The gauge bosons corresponding to the
  broken gauge symmetries obtain their masses
  not through the Higgs mechanism but
  as the Kaluza-Klein (KK) states\cite{KK}
  with masses of
  $n/R$, where $R$ is the compactification
  scale with the positive integer $n$.
It has been shown in Refs.\cite{Sekhar1,Sekhar3,abe} that
  higher dimensional gauge theories
  preserve the
  unitarity in the sense that the power
  law behavior $O(E^4/m^4)$ or $O(E^2/m^2)$ is canceled.
The related discussion are shown
  in Refs.\cite{1, 2, 3, uni, DeCurtis, Hall:2001tn, Schwartz, Murayama}.

In the previous paper\cite{abe}
   we have studied the unitarity bounds of the
   extra dimensional gauge theory
   where the gauge symmetry is broken
   by the nontrivial boundary conditions.
We have calculated the amplitudes of the process
   including four massive gauge bosons in the external lines
   in the framework of the 5-dimensional
   SM and the $SU(5)$ GUT
   whose 5th-dimensional coordinate
   is compactified on $S^1/Z_2$.
We have shown that the power behavior of
    $O(E^4/m^4)$ and that of $O(E^2/m^2)$
   in the amplitude  both vanish,
   and that the broken gauge theory through
   the orbifolding preserves unitarity at high energy.
It has been noted that
   the structure of the interactions
   among KK states
   are crucial for conserving the unitarity.
The calculations have been done
   in the unitary gauge.
The fifth gauge field was
   {\it gauged away} and absorbed into the
   longitudinal component of the 4D gauge field
   through the appropriate gauge transformation
   compatible with the boundary
   conditions on the $S^1/Z_2$ orbifold.

In this paper
   we reexamine the extra dimensional gauge theory
   where the gauge symmetry is broken
   by the nontrivial boundary conditions
We show that the 5th gauge field is
   a {\it would-be NG-like} field and 
   derive the 4D equivalence theorem
   in the 't Hooft-Feynman gauge.
It is organized as follows.
In section 2 we discuss
   the gauge transformation
   compatible with the boundary
   condition on the $S^1/Z_2$ orbifold.
In section 3 we give the BRST formalism on the 4D theory
   which is obtained through the integration of the 5D theory.
In section 4 we derive the 4D equivalence theorem
   and note that the 5th gauge field is
   the would-be NG-like field.
In section 5 we present the amplitudes of the process
   including four massive gauge bosons in the external lines,
   in comparison with the ones for the connected reactions
   where the gauge fields are replaced
   by the corresponding would-be NG-like fields.
The final section is devoted to summary and discussion.

%%%%%%%%%%%%%%%%%%%%%%%%%%%%%%%%%%%%%%%%%%%%%%%%%%%%%%%%%%%%%%%%%%%
%%%%%%%%%%%%%%%%%%%%% SECTION %%%%%%%%%%%%%%%%%%%%%%%%%%%%%%%%%%%%%
%%%%%%%%%%%%%%%%%%%%%%%%%%%%%%%%%%%%%%%%%%%%%%%%%%%%%%%%%%%%%%%%%%%
\section{Gauge symmetry on orbifold}

To start with, we show the setup.
We consider the 5D gauge theory with
   the gauge field living
   in the bulk.
We denote the 5-dimensional coordinate
   as $y$, which is
   compactified on an $S^1/Z_2$
   orbifold.
Under the $Z_2$ parity transformation,
   $y \rightarrow -y$,
   the gauge fields $A_\nu(x^\mu,y)$ $(\nu = 0-3)$
   and $A_5(x^\mu,y)$ transform as
\begin{eqnarray}
\label{1}
A_\nu(x^\mu,y) &\to& A_\nu(x^\mu,-y) = PA_\nu(x^\mu,y)P^{-1},\\
\label{2}
A_5(x^\mu,y) &\to& A_5(x^\mu,-y) = -PA_5(x^\mu,y)P^{-1},
\end{eqnarray}
where $P$ is the operator of
   $Z_2$ transformation.
Two walls at $y=0$ and $\pi R$ are
   fixed points under the $Z_2$ transformation.
The physical space can be taken to be $0 \leq y \leq \pi R$.
Here we take $Z_2$ as $P=1$, so that
   the mode expansion of $A_\nu$ ($A_5$) for the
   5D coordinate is given by series of cosine (sine) functions.
Besides, we consider the nontrivial
   boundary conditions $T: y \rightarrow y+2\pi R$,
   where the parity (reflection)
   operator $P'$ around $y=\pi R$
   is given by $P'=TP$.
On this orbifold, the fields $A_\nu(x^\mu,y)$ and $A_5(x^\mu,y)$
   are divided into
\begin{eqnarray}
\label{3}
A_{\nu +}(x^\mu,y) &=& \sum_{n=0}^{\infty}
   \frac{1}{\sqrt{2^{\delta_{n,0}}\pi R}}\; A_{\nu}^{(n)}
            (x^\mu)\; \cos \frac{ny}{R},\\
\label{4}
A_{\nu -}(x^\mu,y) &=& \sum_{n=0}^{\infty}
    \frac{1}{\sqrt{\pi R}}\; A_{\nu}^{(n+{1\over2})}
            (x^\mu)\; \cos \frac{(n+{1\over2})y}{R}, \\
\label{5}
A_{5 +}(x^\mu,y) &=& \sum_{n=0}^{\infty}
    \frac{1}{\sqrt{\pi R}}\; A_{5}^{(n+{1\over2})}
            (x^\mu)\; \sin \frac{(n+{1\over2})y}{R}, \\
\label{6}
A_{5 -}(x^\mu,y) &=& \sum_{n=0}^{\infty}
   \frac{1}{\sqrt{\pi R}}\; A_{5}^{(n+1)}
            (x^\mu)\; \sin \frac{(n+1)y}{R},
\end{eqnarray}
according to the eigenvalues, $\pm1$,
   of the parity $P'$.

The gauge symmetry
   is broken by  the nontrivial
   parity operator $P'$ in the gauge group basis.
In the case of the 5D $SU(3)_W$ theory,
   the $Z_2$ parity operator, $P'= diag.(1,1,-1)$,
   realizes the gauge reduction of
   $SU(3)_W \rightarrow SU(2)_L\times U(1)_Y$\cite{Hall:2001zb}.
In case of the 5D $SU(5)$ theory,
   the $Z_2$ parity operator, $P'= diag.(-1,-1,-1,1,1)$,
   realizes the gauge reduction of
   $SU(5) \rightarrow SU(3)_c \times SU(2)_L\times U(1)_Y$\cite{5d}.

Hereafter, $a$ ($\hat{a}$) denotes
   {\it unbroken} ({\it broken}) gauge-indeces.
The gauge fields are expanded as
\begin{eqnarray}
\label{7}
A_{\nu}^{a}(x^\mu,y) &=& \sum_{n=0}^{\infty}
   \frac{1}{\sqrt{2^{\delta_{n,0}}\pi R}}\; A_{\nu}^{a(n)}
            (x^\mu)\; \cos \frac{ny}{R},\\
\label{8}
A_{\nu}^{\hat{a}}(x^\mu,y) &=& \sum_{n=0}^{\infty}
    \frac{1}{\sqrt{\pi R}}\; A_{\nu}^{\hat{a}(n+{1\over2})}
            (x^\mu)\; \cos \frac{(n+{1\over2})y}{R}, \\
\label{9}
A_{5}^{\hat{a}}(x^\mu,y) &=& \sum_{n=0}^{\infty}
    \frac{1}{\sqrt{\pi R}}\; A_{5}^{\hat{a}(n+{1\over2})}
            (x^\mu)\; \sin \frac{(n+{1\over2})y}{R}, \\
\label{10}
A_{5}^{a}(x^\mu,y) &=& \sum_{n=0}^{\infty}
   \frac{1}{\sqrt{\pi R}}\; A_{5}^{a(n+1)}
            (x^\mu)\; \sin \frac{(n+1)y}{R}.
\end{eqnarray}

The 5D Lagrangian on the orbifold is given by
\begin{eqnarray}
\label{11}
{\cal L}_{5} &=&
   - {1\over4} (F_{\mu\nu}^{a})^{2}
   - {1\over4} (F_{\mu\nu}^{\hat{a}})^{2}
   - {1\over2} (F_{\mu5}^{\hat{a}})^{2}
   - {1\over2} (F_{\mu5}^{a})^{2}.
\end{eqnarray}
Here,
\begin{eqnarray}
\label{12}
F_{\mu\nu}^{a} &=&
   \partial_\mu A_{\nu}^{a}
   - \partial_\nu A_{\mu}^{a}
   - g_5 (f^{abc} A_{\mu}^{b} A_{\nu}^{c}
   + f^{a\hat{b}\hat{c}} A_{\mu}^{\hat{b}} A_{\nu}^{\hat{c}}),\\
\label{13}
F_{\mu\nu}^{\hat{a}} &=&
   \partial_\mu A_{\nu}^{\hat{a}}
   - \partial_\nu A_{\mu}^{\hat{a}}
   - g_5 (f^{\hat{a}\hat{b}c} A_{\mu}^{\hat{b}} A_{\nu}^{c}
   + f^{\hat{a}b\hat{c}} A_{\mu}^{b} A_{\nu}^{\hat{c}}),\\
\label{14}
F_{\mu5}^{\hat{a}} &=&
   \partial_\mu A_{5}^{\hat{a}}
   - \partial_5 A_{\mu}^{\hat{a}}
   - g_5 (f^{\hat{a}\hat{b}c} A_{\mu}^{\hat{b}} A_{5}^{c}
   + f^{\hat{a}{b}\hat{c}} A_{\mu}^{b} A_{5}^{\hat{c}}),\\
\label{15}
F_{\mu5}^{a} &=&
   \partial_\mu A_{5}^{a}
   - \partial_5 A_{\mu}^{a}
   - g_5 (f^{abc} A_{\mu}^{b} A_{5}^{c}
   + f^{a\hat{b}\hat{c}} A_{\mu}^{\hat{b}} A_{5}^{\hat{c}}),
\end{eqnarray}
where $g_5$ is the 5D gauge coupling,
   $f^{abc}$, $f^{a\hat{b}\hat{c}}$, $f^{\hat{a}\hat{b}c}$,
   $f^{\hat{a}{b}\hat{c}}$ the structure constant,
   and $A_{\nu}^{a}$, $A_{\nu}^{\hat{a}}$, $A_{5}^{\hat{a}}$,
   $A_{5}^{a}$ in Eqs.(\ref{7})--(\ref{10}).

The above 5D Lagrangian  is invariant under
   the 5D gauge transformation as
\begin{eqnarray}
\label{16}
\delta A_{\mu}^{a} &=&
   \partial_\mu \epsilon ^{a}
   - g_5 (f^{abc} A_{\mu}^{b} \epsilon ^{c}
   + f^{a\hat{b}\hat{c}} A_{\mu}^{\hat{b}} \epsilon ^{\hat{c}}),\\
\label{17}
\delta A_{\mu}^{\hat{a}} &=&
   \partial_\mu \epsilon ^{\hat{a}}
   - g_5 (f^{\hat{a}\hat{b}c} A_{\mu}^{\hat{b}} \epsilon ^{c}
   + f^{\hat{a}b\hat{c}} A_{\mu}^{b} \epsilon ^{\hat{c}}),\\
\label{18}
\delta A_{5}^{\hat{a}} &=&
   \partial_5 \epsilon ^{\hat{a}}
   - g_5 (f^{\hat{a}\hat{b}c} A_{5}^{\hat{b}} \epsilon ^{c}
   + f^{\hat{a}{b}\hat{c}} A_{5}^{b} \epsilon ^{\hat{c}}),\\
\label{19}
\delta A_{5}^{a} &=&
   \partial_5 \epsilon ^{a}
   - g_5 (f^{abc} A_{5}^{b} \epsilon ^{c}
   + f^{a\hat{b}\hat{c}} A_{5}^{\hat{b}} \epsilon ^{\hat{c}}),
\end{eqnarray}
where the 5D gauge functions $\epsilon^{a}(x^\mu,y)$ and
$\epsilon^{\hat{a}}(x^\mu,y)$ are given as
\begin{eqnarray}
\label{20}
\epsilon^{a}(x^\mu,y) &=& \sum_{n=0}^{\infty}
   \frac{1}{\sqrt{2^{\delta_{n,0}}\pi R}}\; \epsilon^{a(n)}
            (x^\mu)\; \cos \frac{ny}{R},\\
\label{21}
\epsilon^{\hat{a}}(x^\mu,y) &=& \sum_{n=0}^{\infty}
    \frac{1}{\sqrt{\pi R}}\; \epsilon^{\hat{a}(n+{1\over2})}
            (x^\mu)\; \cos \frac{(n+{1\over2})y}{R}.
\end{eqnarray}
It is noted that the above transformation is compatible
   with the boundary conditions on the orbifold.

The above 5D gauge transformation can be rewritten
in terms of the relevant KK modes as
\begin{eqnarray}
\label{22}
\delta A_{\mu}^{a(0)} &=&
   \partial_\mu \epsilon ^{a(0)}
   - g_4 \sum_{m=1}^{\infty}
   \left[
   f^{abc} A_{\mu}^{b(m-1)} \epsilon ^{c(m-1)}
   + f^{a\hat{b}\hat{c}} A_{\mu}^{\hat{b}(m-{1\over2})} \epsilon
^{\hat{c}(m-{1\over2})}
   \right],\\
\label{23}
\delta A_{\mu}^{a(n)} &=&
   \partial_\mu \epsilon ^{a(n)}
   - g_4 f^{abc}
   \left[
   A_{\mu}^{b(0)} \epsilon ^{c(n)}
   + A_{\mu}^{b(n)} \epsilon ^{c(0)}
   \right]
   \nonumber\\
   &&
   - \frac{g_4}{\sqrt{2}}\; f^{abc}
   \left[
   \sum_{m=1}^{n-1} A_{\mu}^{b(n-m)} \epsilon ^{c(m)}
   + \sum_{m=1}^{\infty} A_{\mu}^{b(n+m)} \epsilon ^{c(m)}
   \right.\nonumber\\ &&\hspace{6cm}\left.
   + \sum_{m=n+1}^{\infty} A_{\mu}^{b(m-n)} \epsilon ^{c(m)}
   \right]
   \nonumber\\
   &&
   - \frac{g_4}{\sqrt{2}}\; f^{a\hat{b}\hat{c}}
   \left[
   \sum_{m=1}^{n} A_{\mu}^{\hat{b}(n-m+{1\over2})} \epsilon
^{\hat{c}(m-{1\over2})}
   + \sum_{m=1}^{\infty} A_{\mu}^{\hat{b}(n+m-{1\over2})} \epsilon
^{\hat{c}(m-{1\over2})}
   \right.\nonumber\\ &&\hspace{6cm}\left.
   + \sum_{m=n+1}^{\infty} A_{\mu}^{\hat{b}(m-n-{1\over2})} \epsilon
^{\hat{c}(m-{1\over2})}
   \right],\\
\label{24}
\delta A_{\mu}^{\hat{a}(n-{1\over2})} &=&
   \partial_\mu \epsilon ^{\hat{a}(n-{1\over2})}
   - g_4
   \left[
   f^{\hat{a}\hat{b}c} A_{\mu}^{\hat{b}(n-{1\over2})} \epsilon ^{c(0)}
   + f^{\hat{a}b\hat{c}} A_{\mu}^{b(0)} \epsilon ^{\hat{c}(n-{1\over2})}
   \right]
   \nonumber\\
   &&
   - \frac{g_4}{\sqrt{2}}\; f^{\hat{a}\hat{b}c}
   \left[
   \sum_{m=1}^{n-1} A_{\mu}^{\hat{b}(n-m-{1\over2})} \epsilon ^{c(m)}
   + \sum_{m=1}^{\infty} A_{\mu}^{\hat{b}(n+m-{1\over2})} \epsilon ^{c(m)}
   \right.\nonumber\\ &&\hspace{6cm}\left.
   + \sum_{m=n}^{\infty} A_{\mu}^{\hat{b}(m-n+{1\over2})} \epsilon ^{c(m)}
   \right]
   \nonumber\\
   &&
   - \frac{g_4}{\sqrt{2}}\; f^{\hat{a}b\hat{c}}
   \left[
   \sum_{m=1}^{n-1} A_{\mu}^{b(n-m)} \epsilon ^{\hat{c}(m-{1\over2})}
   + \sum_{m=1}^{\infty} A_{\mu}^{b(n+m-1)} \epsilon ^{\hat{c}(m-{1\over2})}
   \right.\nonumber\\ &&\hspace{6cm}\left.
   + \sum_{m=n+1}^{\infty} A_{\mu}^{b(m-n)} \epsilon ^{\hat{c}(m-{1\over2})}
   \right],\\
\label{25}
\delta A_{5}^{\hat{a}(n-{1\over2})} &=&
   - \frac{n-{1\over2}}{R} \epsilon ^{\hat{a}(n-{1\over2})}
   - g_4 f^{\hat{a}\hat{b}c} A_{5}^{\hat{b}(n-{1\over2})} \epsilon ^{c(0)}
   \nonumber\\
   &&
   - \frac{g_4}{\sqrt{2}}\; f^{\hat{a}\hat{b}c}
   \left[
   \sum_{m=1}^{n-1} A_{5}^{\hat{b}(n-m-{1\over2})} \epsilon ^{c(m)}
   + \sum_{m=1}^{\infty} A_{5}^{\hat{b}(n+m-{1\over2})} \epsilon ^{c(m)}
   \right.\nonumber\\ &&\hspace{6cm}\left.
   - \sum_{m=n}^{\infty} A_{5}^{\hat{b}(m-n+{1\over2})} \epsilon ^{c(m)}
   \right]
   \nonumber\\
   &&
   - \frac{g_4}{\sqrt{2}}\; f^{\hat{a}b\hat{c}}
   \left[
   \sum_{m=1}^{n-1} A_{5}^{b(n-m)} \epsilon ^{\hat{c}(m-{1\over2})}
   + \sum_{m=1}^{\infty} A_{5}^{b(n+m-1)} \epsilon ^{\hat{c}(m-{1\over2})}
   \right.\nonumber\\ &&\hspace{6cm}\left.
   - \sum_{m=n+1}^{\infty} A_{5}^{b(m-n)} \epsilon ^{\hat{c}(m-{1\over2})}
   \right],\\
\label{26}
\delta A_{5}^{a(n)} &=&
   - \frac{n}{R} \epsilon ^{a(n)}
   - g_4 f^{abc} A_{5}^{b(n)} \epsilon ^{c(0)}
   \nonumber\\
   &&
   - \frac{g_4}{\sqrt{2}}\; f^{abc}
   \left[
   \sum_{m=1}^{n-1} A_{5}^{b(n-m)} \epsilon ^{c(m)}
   + \sum_{m=1}^{\infty} A_{5}^{b(n+m)} \epsilon ^{c(m)}
   \right.\nonumber\\ &&\hspace{6cm}\left.
   - \sum_{m=n+1}^{\infty} A_{5}^{b(m-n)} \epsilon ^{c(m)}
   \right]
   \nonumber\\
   &&
   - \frac{g_4}{\sqrt{2}}\; f^{a\hat{b}\hat{c}}
   \left[
   \sum_{m=1}^{n} A_{5}^{\hat{b}(n-m+{1\over2})} \epsilon 
^{\hat{c}(m-{1\over2})}
   + \sum_{m=1}^{\infty} A_{5}^{\hat{b}(n+m-{1\over2})} \epsilon
^{\hat{c}(m-{1\over2})}
   \right.\nonumber\\ &&\hspace{6cm}\left.
   - \sum_{m=n+1}^{\infty} A_{5}^{\hat{b}(m-n-{1\over2})} \epsilon
^{\hat{c}(m-{1\over2})}
   \right],
\end{eqnarray}
where $n$ and $m$ are positive integers
   and $g_4$ is the 4D gauge coupling
   which is related to the 5D gauge coupling, $g_5$,
   as $g_4=g_5/\sqrt{2\pi R}$.  .

%%%%%%%%%%%%%%%%%%%%%%%%%%%%%%%%%%%%%%%%%%%%%%%%%%%%%%%%%%%%%%%%%%%
%%%%%%%%%%%%%%%%%%%%% SECTION %%%%%%%%%%%%%%%%%%%%%%%%%%%%%%%%%%%%%
%%%%%%%%%%%%%%%%%%%%%%%%%%%%%%%%%%%%%%%%%%%%%%%%%%%%%%%%%%%%%%%%%%%
\section{4D Lagrangian and BRST formalism}

The 4D Lagrangian is obtained
   by integrating the 5D Lagrangian
   and devided into the kinetic term ${\cal L}_{KE}$
   and the interaction term ${\cal L}_{INT}$
\begin{eqnarray}
\label{27}
{\cal L}_{4} =
   \int_{0}^{2\pi R} dy {\cal L}_{5} =
   {\cal L}_{KE}+{\cal L}_{INT}.
\end{eqnarray}
This Lagrangian is invariant under the 4D gauge transformation
   as shown in Eqs.(\ref{22})--(\ref{26}).
The kinetic term ${\cal L}_{KE}$ is given as
\begin{eqnarray}
\label{28}
{\cal L}_{KE} &=&
   - {1\over4}
   \left[
   \partial_\mu A_{\nu}^{a(0)} - \partial_\nu A_{\mu}^{a(0)}
   \right]^{2}
   \nonumber\\
   &&
   - {1\over4} \sum_{n=1}^{\infty}
   \left[
   \partial_\mu A_{\nu}^{a(n)} - \partial_\nu A_{\mu}^{a(n)}
   \right]^{2}
   - {1\over2} \sum_{n=1}^{\infty}
   \left[
   \partial_\mu A_{5}^{a(n)} + M_{n} A_{\mu}^{a(n)}
   \right]^{2}
   \nonumber\\
   &&
   - {1\over4} \sum_{n=1}^{\infty}
   \left[
   \partial_\mu A_{\nu}^{\hat{a}(n-{1\over2})}
   - \partial_\nu A_{\mu}^{\hat{a}(n-{1\over2})}
   \right]^{2}
   \nonumber\\ &&\hspace{4cm}
   - {1\over2} \sum_{n=1}^{\infty}
   \left[
   \partial_\mu A_{5}^{\hat{a}(n-{1\over2})}
   + M_{n-{1\over2}} A_{\mu}^{\hat{a}(n-{1\over2})}
   \right]^{2},
\end{eqnarray}
where $M_{n} = n/R$ and $M_{n-{1\over2}} = (n-{1\over2})/R$
   are the masses of the KK vector bosons.

We impose an $R_{\xi}$ gauge-fixing of the form
\begin{eqnarray}
\label{29}
{\cal L}_{GF} =
   - \frac{1}{2{\xi}_0}\;
   \left[ F^{a(0)} \right]^{2}
   - \sum_{n=1}^{\infty} \frac{1}{2{\xi}_n}\;
   \left[ F^{a(n)} \right]^{2}
   - \sum_{n=1}^{\infty} \frac{1}{2{\xi}_{n-{1\over2}}}\;
   \left[ F^{\hat{a}(n-{1\over2})} \right]^{2},
\end{eqnarray}
where ${\xi}_0$, ${\xi}_n$ and ${\xi}_{n-{1\over2}}$
   are arbitrary gauge parameters and
\begin{eqnarray}
\label{30}
F^{a(0)} &=&
   \partial^\mu A_{\mu}^{a(0)},\\
\label{31}
F^{a(n)} &=&
   \partial^\mu A_{\mu}^{a(n)}
   + {\xi}_n M_{n} A_{5}^{a(n)},\\
\label{32}
F^{\hat{a}(n-{1\over2})} &=&
   \partial^\mu A_{\mu}^{\hat{a}(n-{1\over2})}
   + {\xi}_{n-{1\over2}} M_{n-{1\over2}} A_{5}^{\hat{a}(n-{1\over2})}.
\end{eqnarray}
The above gauge-fixing term eliminates
   the kinetic term mixing between
   $A_{\mu}^{a(n)}$ ($A_{\mu}^{\hat{a}(n-{1\over2})}$) and
   $A_{5}^{a(n)}$ ($A_{5}^{\hat{a}(n-{1\over2})}$),
   and we may identify the
   $A_{5}^{a(n)}$ and $A_{5}^{\hat{a}(n-{1\over2})}$ modes
   as the {\it would-be NG-like} fields.
The $A_{5}^{a(n)}$ and $A_{5}^{\hat{a}(n-{1\over2})}$
   have gauge-dependent masses
   $M_{5(n)}^2 = {\xi}_n M_{n}^2$ and
   $M_{5(n-{1\over2})}^2 = {\xi}_{n-{1\over2}} M_{n-{1\over2}}^2$, respectively.
In the previous paper\cite{abe}
   we have taken the unitary gauge,
   ${\xi}_n = {\infty}$ and
   ${\xi}_{n-{1\over2}} = {\infty}$,
   where the $A_{5}^{a(n)}$ and $A_{5}^{\hat{a}(n-{1\over2})}$ decouple since
   $M_{5(n)} \rightarrow {\infty}$ and
   $M_{5(n-{1\over2})} \rightarrow {\infty}$.

The appropriate Faddeev-Popov ghost term is 
\begin{eqnarray}
\label{33}
{\cal L}_{FP} =
   - \sum_{n=0}^{\infty}
   {\bar{\eta}}^{a(n)} {\delta}_B F^{a(n)}
   - \sum_{n=1}^{\infty}
   {\bar{\eta}}^{\hat{a}(n-{1\over2})} {\delta}_B F^{\hat{a}(n-{1\over2})},
\end{eqnarray}
where the BRST transformation is given as follows
\begin{eqnarray}
\label{34}
\delta_B {\bar{\eta}}^{a(n)} &=&
   - \frac{1}{{\xi}_n}\; F^{a(n)},\\
\label{35}
\delta_B {\bar{\eta}}^{\hat{a}(n-{1\over2})} &=&
   - \frac{1}{{\xi}_{n-{1\over2}}}\; F^{\hat{a}(n-{1\over2})},\\
\label{36}
\delta_B A_{\mu}^{a(0)} &=&
   \partial_\mu \eta ^{a(0)}
   - g_4 f^{abc} A_{\mu}^{b(0)} \eta ^{c(0)}
   \nonumber\\
   &&
   - g_4 \sum_{m=1}^{\infty}
   \left[
   f^{abc} A_{\mu}^{b(m)} \eta ^{c(m)}
   + f^{a\hat{b}\hat{c}} A_{\mu}^{\hat{b}(m-{1\over2})} \eta
^{\hat{c}(m-{1\over2})}
   \right],\\
\label{37}
\delta_B A_{\mu}^{a(n)} &=&
   \partial_\mu \eta ^{a(n)}
   - g_4 f^{abc}
   \left[
   A_{\mu}^{b(0)} \eta ^{c(n)}
   + A_{\mu}^{b(n)} \eta ^{c(0)}
   \right]
   \nonumber\\
   &&
   - \frac{g_4}{\sqrt{2}}\; f^{abc}
   \left[
   \sum_{m=1}^{n-1} A_{\mu}^{b(n-m)} \eta ^{c(m)}
   + \sum_{m=1}^{\infty} A_{\mu}^{b(n+m)} \eta ^{c(m)}
   \right.
   \nonumber\\
   &&
   \hspace{5cm}
   \left.
   + \sum_{m=n+1}^{\infty} A_{\mu}^{b(m-n)} \eta ^{c(m)}
   \right]
   \nonumber\\
   &&
   - \frac{g_4}{\sqrt{2}}\; f^{a\hat{b}\hat{c}}
   \left[
   \sum_{m=1}^{n} A_{\mu}^{\hat{b}(n-m+{1\over2})} \eta ^{\hat{c}(m-{1\over2})}
   + \sum_{m=1}^{\infty} A_{\mu}^{\hat{b}(n+m-{1\over2})} \eta
^{\hat{c}(m-{1\over2})}
   \right.
   \nonumber\\
   &&
   \hspace{5cm}
   \left.
   + \sum_{m=n+1}^{\infty} A_{\mu}^{\hat{b}(m-n-{1\over2})} \eta
^{\hat{c}(m-{1\over2})}
   \right],\\
\label{38}
\delta_B A_{\mu}^{\hat{a}(n-{1\over2})} &=&
   \partial_\mu \eta ^{\hat{a}(n-{1\over2})}
   - g_4
   \left[
   f^{\hat{a}\hat{b}c} A_{\mu}^{\hat{b}(n-{1\over2})} \eta ^{c(0)}
   + f^{\hat{a}b\hat{c}} A_{\mu}^{b(0)} \eta ^{\hat{c}(n-{1\over2})}
   \right]
   \nonumber\\
   &&
   - \frac{g_4}{\sqrt{2}}\; f^{\hat{a}\hat{b}c}
   \left[
   \sum_{m=1}^{n-1} A_{\mu}^{\hat{b}(n-m-{1\over2})} \eta ^{c(m)}
   + \sum_{m=1}^{\infty} A_{\mu}^{\hat{b}(n+m-{1\over2})} \eta ^{c(m)}
   \right.
   \nonumber\\
   &&
   \hspace{5cm}
   \left.
   + \sum_{m=n}^{\infty} A_{\mu}^{\hat{b}(m-n+{1\over2})} \eta ^{c(m)}
   \right]
   \nonumber\\
   &&
   - \frac{g_4}{\sqrt{2}}\; f^{\hat{a}b\hat{c}}
   \left[
   \sum_{m=1}^{n-1} A_{\mu}^{b(n-m)} \eta ^{\hat{c}(m-{1\over2})}
   + \sum_{m=1}^{\infty} A_{\mu}^{b(n+m-1)} \eta ^{\hat{c}(m-{1\over2})}
   \right.
   \nonumber\\
   &&
   \hspace{5cm}
   \left.
   + \sum_{m=n+1}^{\infty} A_{\mu}^{b(m-n)} \eta ^{\hat{c}(m-{1\over2})}
   \right],\\
\label{39}
\delta_B A_{5}^{\hat{a}(n-{1\over2})} &=&
   - M_{n-{1\over2}} \eta ^{\hat{a}(n-{1\over2})}
   - g_4 f^{\hat{a}\hat{b}c} A_{5}^{\hat{b}(n-{1\over2})} \eta ^{c(0)}
   \nonumber\\
   &&
   - \frac{g_4}{\sqrt{2}}\; f^{\hat{a}\hat{b}c}
   \left[
   \sum_{m=1}^{n-1} A_{5}^{\hat{b}(n-m-{1\over2})} \eta ^{c(m)}
   + \sum_{m=1}^{\infty} A_{5}^{\hat{b}(n+m-{1\over2})} \eta ^{c(m)}
   \right.
   \nonumber\\
   &&
   \hspace{5cm}
   \left.
   - \sum_{m=n}^{\infty} A_{5}^{\hat{b}(m-n+{1\over2})} \eta ^{c(m)}
   \right]
   \nonumber\\
   &&
   - \frac{g_4}{\sqrt{2}}\; f^{\hat{a}b\hat{c}}
   \left[
   \sum_{m=1}^{n-1} A_{5}^{b(n-m)} \eta ^{\hat{c}(m-{1\over2})}
   + \sum_{m=1}^{\infty} A_{5}^{b(n+m-1)} \eta ^{\hat{c}(m-{1\over2})}
   \right.
   \nonumber\\
   &&
   \hspace{5cm}
   \left.
   - \sum_{m=n+1}^{\infty} A_{5}^{b(m-n)} \eta ^{\hat{c}(m-{1\over2})}
   \right],\\
\label{40}
\delta_B A_{5}^{a(n)} &=&
   - M_{n} \eta ^{a(n)}
   - g_4 f^{abc} A_{5}^{b(n)} \eta ^{c(0)}
   \nonumber\\
   &&
   - \frac{g_4}{\sqrt{2}}\; f^{abc}
   \left[
   \sum_{m=1}^{n-1} A_{5}^{b(n-m)} \eta ^{c(m)}
   + \sum_{m=1}^{\infty} A_{5}^{b(n+m)} \eta ^{c(m)}
   \right.
   \nonumber\\
   &&\hspace{5cm}
   \left.
   - \sum_{m=n+1}^{\infty} A_{5}^{b(m-n)} \eta ^{c(m)}
   \right]
   \nonumber\\
   &&
   - \frac{g_4}{\sqrt{2}}\; f^{a\hat{b}\hat{c}}
   \left[
   \sum_{m=1}^{n} A_{5}^{\hat{b}(n-m+{1\over2})} \eta ^{\hat{c}(m-{1\over2})}
   + \sum_{m=1}^{\infty} A_{5}^{\hat{b}(n+m-{1\over2})} \eta 
^{\hat{c}(m-{1\over2})}
   \right.
   \nonumber\\
   &&
   \hspace{5cm}
   \left.
   - \sum_{m=n+1}^{\infty} A_{5}^{\hat{b}(m-n-{1\over2})} \eta
^{\hat{c}(m-{1\over2})}
   \right].
\end{eqnarray}
Inserting the above relations {\it etc}. to Eq.(33),
   we can express the Faddeev-Popov Lagrangian
   in terms of the ghosts ${\eta}$ and
   the anti-ghosts ${\bar{\eta}}$ {\it etc}. such as
\begin{eqnarray}
\label{41}
{\cal L}_{FP} &=& \partial^\mu {\bar{\eta}}^{a(0)}
   \left[
   \partial_\mu \eta ^{a(0)}
   - g_4 f^{abc} A_{\mu}^{b(0)} \eta ^{c(0)}
   \right.
   \nonumber\\
   &&
   \left.
   - g_4 \sum_{m=1}^{\infty}
   \left[
   f^{abc} A_{\mu}^{b(m)} \eta ^{c(m)}
   + f^{a\hat{b}\hat{c}} A_{\mu}^{\hat{b}(m-{1\over2})} \eta
^{\hat{c}(m-{1\over2})}
   \right]
   \right]
   \nonumber\\
   &&
   + \sum_{n=1}^{\infty} \partial^\mu {\bar{\eta}}^{a(n)}
   \left[
   \partial_\mu \eta ^{a(n)}
   - g_4 f^{abc}
   \left[
   A_{\mu}^{b(0)} \eta ^{c(n)}
   + A_{\mu}^{b(n)} \eta ^{c(0)}
   \right]
   \right.
   \nonumber\\
   &&
   \left.
   - \frac{g_4}{\sqrt{2}}\; f^{abc}
   \left[
   \sum_{m=1}^{n-1} A_{\mu}^{b(n-m)} \eta ^{c(m)}
   + \sum_{m=1}^{\infty} A_{\mu}^{b(n+m)} \eta ^{c(m)}
   \right.
   \right.
   \nonumber\\
   &&
   \hspace{5cm}
   \left.
   + \sum_{m=n+1}^{\infty} A_{\mu}^{b(m-n)} \eta ^{c(m)}
   \right]
   \nonumber\\
   &&
   - \frac{g_4}{\sqrt{2}}\; f^{a\hat{b}\hat{c}}
   \left[
   \sum_{m=1}^{n} A_{\mu}^{\hat{b}(n-m+{1\over2})} \eta ^{\hat{c}(m-{1\over2})}
   + \sum_{m=1}^{\infty} A_{\mu}^{\hat{b}(n+m-{1\over2})} \eta
^{\hat{c}(m-{1\over2})}
   \right.
   \nonumber\\
   &&
   \hspace{5cm}
   \left.
   \left.
   + \sum_{m=n+1}^{\infty} A_{\mu}^{\hat{b}(m-n-{1\over2})} \eta
^{\hat{c}(m-{1\over2})}
   \right]
   \right]
   \nonumber\\
   &&
   + \sum_{n=1}^{\infty} \partial^\mu {\bar{\eta}}^{\hat{a}(n-{1\over2})}
   \left[
   \partial_\mu \eta ^{\hat{a}(n-{1\over2})}
   - g_4
   \left[
   f^{\hat{a}\hat{b}c} A_{\mu}^{\hat{b}(n-{1\over2})} \eta ^{c(0)}
   + f^{\hat{a}b\hat{c}} A_{\mu}^{b(0)} \eta ^{\hat{c}(n-{1\over2})}
   \right]
   \right.
   \nonumber\\
   &&
   - \frac{g_4}{\sqrt{2}}\; f^{\hat{a}\hat{b}c}
   \left[
   \sum_{m=1}^{n-1} A_{\mu}^{\hat{b}(n-m-{1\over2})} \eta ^{c(m)}
   + \sum_{m=1}^{\infty} A_{\mu}^{\hat{b}(n+m-{1\over2})} \eta ^{c(m)}
   \right.
   \nonumber\\
   &&
   \hspace{5cm}
   \left.
   + \sum_{m=n}^{\infty} A_{\mu}^{\hat{b}(m-n+{1\over2})} \eta ^{c(m)}
   \right]
   \nonumber\\
   &&
   - \frac{g_4}{\sqrt{2}}\; f^{\hat{a}b\hat{c}}
   \left[
   \sum_{m=1}^{n-1} A_{\mu}^{b(n-m)} \eta ^{\hat{c}(m-{1\over2})}
   + \sum_{m=1}^{\infty} A_{\mu}^{b(n+m-1)} \eta ^{\hat{c}(m-{1\over2})}
   \right.
   \nonumber\\
   &&
   \hspace{5cm}
   \left.
   \left.
   + \sum_{m=n+1}^{\infty} A_{\mu}^{b(m-n)} \eta ^{\hat{c}(m-{1\over2})}
   \right]
   \right]
   \nonumber\\
   &&
   - \sum_{n=1}^{\infty} \xi_{n-{1\over2}} M_{n-{1\over2}}
{\bar{\eta}}^{\hat{a}(n-{1\over2})}
   \left[
   - M_{n-{1\over2}} \eta ^{\hat{a}(n-{1\over2})}
   - g_4 f^{\hat{a}\hat{b}c} A_{5}^{\hat{b}(n-{1\over2})} \eta ^{c(0)}
   \right.
   \nonumber\\
   &&
   - \frac{g_4}{\sqrt{2}}\; f^{\hat{a}\hat{b}c}
   \left[
   \sum_{m=1}^{n-1} A_{5}^{\hat{b}(n-m-{1\over2})} \eta ^{c(m)}
   + \sum_{m=1}^{\infty} A_{5}^{\hat{b}(n+m-{1\over2})} \eta ^{c(m)}
   \right.
   \nonumber\\
   &&
   \hspace{5cm}
   \left.
   - \sum_{m=n}^{\infty} A_{5}^{\hat{b}(m-n+{1\over2})} \eta ^{c(m)}
   \right]
   \nonumber\\
   &&
   - \frac{g_4}{\sqrt{2}}\; f^{\hat{a}b\hat{c}}
   \left[
   \sum_{m=1}^{n-1} A_{5}^{b(n-m)} \eta ^{\hat{c}(m-{1\over2})}
   + \sum_{m=1}^{\infty} A_{5}^{b(n+m-1)} \eta ^{\hat{c}(m-{1\over2})}
   \right.
   \nonumber\\
   &&
   \hspace{5cm}
   \left.
   \left.
   - \sum_{m=n+1}^{\infty} A_{5}^{b(m-n)} \eta ^{\hat{c}(m-{1\over2})}
   \right]
   \right]
   \nonumber\\
   &&
   - \sum_{n=1}^{\infty} \xi_{n} M_{n} {\bar{\eta}}^{a(n)}
   \left[
   - M_{n} \eta ^{a(n)}
   - g_4 f^{abc} A_{5}^{b(n)} \eta ^{c(0)}
   \right.
   \nonumber\\
   &&
   - \frac{g_4}{\sqrt{2}}\; f^{abc}
   \left[
   \sum_{m=1}^{n-1} A_{5}^{b(n-m)} \eta ^{c(m)}
   + \sum_{m=1}^{\infty} A_{5}^{b(n+m)} \eta ^{c(m)}
   \right.
   \nonumber\\
   &&\hspace{5cm}
   \left.
   - \sum_{m=n+1}^{\infty} A_{5}^{b(m-n)} \eta ^{c(m)}
   \right]
   \nonumber\\
   &&
   - \frac{g_4}{\sqrt{2}}\; f^{a\hat{b}\hat{c}}
   \left[
   \sum_{m=1}^{n} A_{5}^{\hat{b}(n-m+{1\over2})} \eta ^{\hat{c}(m-{1\over2})}
   + \sum_{m=1}^{\infty} A_{5}^{\hat{b}(n+m-{1\over2})} \eta 
^{\hat{c}(m-{1\over2})}
   \right.
   \nonumber\\
   &&
   \hspace{5cm}
   \left.
   \left.
   - \sum_{m=n+1}^{\infty} A_{5}^{\hat{b}(m-n-{1\over2})} \eta
^{\hat{c}(m-{1\over2})}
   \right]
   \right].
\end{eqnarray}
%

%%%%%%%%%%%%%%%%%%%%%%%%%%%%%%%%%%%%%%%%%%%%%%%%%%%%%%%%%%%%%%%%%%%
%%%%%%%%%%%%%%%%%%%%% SECTION %%%%%%%%%%%%%%%%%%%%%%%%%%%%%%%%%%%%%
%%%%%%%%%%%%%%%%%%%%%%%%%%%%%%%%%%%%%%%%%%%%%%%%%%%%%%%%%%%%%%%%%%%
\section{4D Equivalence theorem}

The equivalence theorem is the statement
  that the scattering amplitede of massive gauge bosons is equal,
  up to some constant factor,
  to that of the corresponding would-be NG-like fields
  in high-enrgy limit.
The latter amplitude behaves as $O(1)$.
Thus unitarity is automatically preserved.
In this section we show
  that this equvalence theorem comes from
  the 4D gauge invariance of the theory.

First we note that the physical state condition on the scattering states $| 
C, in>$, $| B, out>$, $| C', in>$ and $| B', out>$ leads to the equations
\begin{eqnarray}
\label{42}
< B, out | \delta_B {\bar{\eta}}^{a(n)} | C, in> = 0,\\
\label{43}
< B', out | \delta_B {\bar{\eta}}^{\hat{a}(n-{1\over2})} | C', in> = 0.
\end{eqnarray}
 From these equations and the BRST transformations Eqs.(34) and (35),
we obtain the Slavnov-Taylor identities as
\begin{eqnarray}
\label{44}
< B, out | F^{a(n)} | C, in> = 0,\\
\label{45}
< B', out | F^{\hat{a}(n-{1\over2})} | C', in> = 0,
\end{eqnarray}
and then
\begin{eqnarray}
\label{46}
< B, out | F^{a(n)} | C, in>_{con} = 0,\\
\label{47}
< B', out | F^{\hat{a}(n-{1\over2})} | C', in>_{con} = 0.
\end{eqnarray}
Adopting the 't Hooft-Feynman gauge
\begin{eqnarray}
\label{48}
{\xi}_{n} = {\xi}_{n-{1\over2}} = 1
\end{eqnarray}
in the gauge-fixing terms in Eqs.(\ref{31}) and (\ref{32}),
we get the relations as
\begin{eqnarray}
\label{49}
< B, out | \partial^\mu A_{\mu}^{a(n)}
   + M_{n} A_{5}^{a(n)} | C, in>_{con} = 0,\\
\label{50}
< B', out | \partial^\mu A_{\mu}^{\hat{a}(n-{1\over2})}
   + M_{n-{1\over2}} A_{5}^{\hat{a}(n-{1\over2})} | C', in>_{con} = 0.
\end{eqnarray}

 From the above relations in Eqs.(\ref{49}) and (\ref{50}), we get
\begin{eqnarray}
\label{51}
-i \frac{p^{\mu}}{M_{n}}\; {\hat{S}}
   [C {\to} B + A_{\mu}^{a(n)} (p, {\lambda})]
   = S [C {\to} B + A_{5}^{a(n)} (p)],\\
\label{52}
-i \frac{p^{\mu}}{M_{n-{1\over2}}}\; {\hat{S}}
   [C' {\to} B' + A_{\mu}^{\hat{a}(n-{1\over2})} (p, {\lambda})]
   = S [C' {\to} B' + A_{5}^{\hat{a}(n-{1\over2})} (p)],
\end{eqnarray}
where
\begin{eqnarray}
\label{53}
\epsilon^{\mu} (p, {\lambda}) {\hat{S}}
   [C {\to} B + A_{\mu}^{a(n)} (p, {\lambda})]
\end{eqnarray}
and
\begin{eqnarray}
\label{54}
\epsilon^{\mu} (p, {\lambda}) {\hat{S}}
   [C' {\to} B' + A_{\mu}^{\hat{a}(n-{1\over2})} (p, {\lambda})]
\end{eqnarray}
denote the S-matrix elements for the processes
   $C {\to} B + A_{\mu}^{a(n)} (p, {\lambda})$ and
   $C' {\to} B' + A_{\mu}^{\hat{a}(n-{1\over2})} (p, {\lambda})$,
   respectively, whereas
\begin{eqnarray}
\label{55}
S [C {\to} B + A_{5}^{a(n)} (p)]
\end{eqnarray}
and
\begin{eqnarray}
\label{56}
S [C' {\to} B' + A_{5}^{\hat{a}(n-{1\over2})} (p)]
\end{eqnarray}
denote the S-matrix elements for the processes
   $C {\to} B + A_{5}^{a(n)} (p)$ and
   $C' {\to} B' + A_{5}^{\hat{a}(n-{1\over2})} (p)$, respectively.

Since at sufficiently high energies,
   the longitudinal-polarization vectors for the fields
   $A_{\mu}^{a(n)} (p, {\lambda})$ and
   $A_{\mu}^{\hat{a}(n-{1\over2})} (p, {\lambda})$
   are given as
\begin{eqnarray}
\label{57}
\epsilon^{\mu} (p, {\lambda} = L) =
   \frac{p^{\mu}}{M_{n}}\; +
   O \left[  \frac{M_{n}}{E}\;  \right]
\end{eqnarray}
and
\begin{eqnarray}
\label{58}
\epsilon^{\mu} (p, {\lambda} = L) =
   \frac{p^{\mu}}{M_{n-{1\over2}}}\; +
   O \left[  \frac{M_{n-{1\over2}}}{E}\;  \right],
\end{eqnarray}
respectively, we obtain
\begin{eqnarray}
\label{59}
-i S [C {\to} B + A_{\mu}^{a(n)} (p, {\lambda} = L)]
   = S [C {\to} B + A_{5}^{a(n)} (p)] +
   O \left[  \frac{M_{n}}{E}\;  \right]
\end{eqnarray}
and
\begin{eqnarray}
\label{60}
-i S [C' {\to} B' + A_{\mu}^{\hat{a}(n-{1\over2})} (p, {\lambda} = L)]
   = S [C' {\to} B' + A_{5}^{\hat{a}(n-{1\over2})} (p)] +
   O \left[  \frac{M_{n-{1\over2}}}{E}\;  \right],
\end{eqnarray}
respectively.

Furthermore, the identity for positive integers $m$, $l$ and $n$ as
\begin{eqnarray}
\label{61}
0 &=&
  < B'', out | \delta_B T [ {\bar{\eta}}^{a(m)} (x_1) F^{b(l)} (x_2) \cdots
   F^{\hat{a}(n-{1\over2})} (x_j) \cdots ] | C'', in>
   \nonumber\\
   &=&
  < B'', out | T [ \delta_B {\bar{\eta}}^{a(m)} (x_1) F^{b(l)} (x_2) \cdots
   F^{\hat{a}(n-{1\over2})} (x_j) \cdots ] | C'', in>
   \nonumber\\
   &&
  - < B'', out | T [ {\bar{\eta}}^{a(m)} (x_1) \delta_B F^{b(l)} (x_2) \cdots
   F^{\hat{a}(n-{1\over2})} (x_j) \cdots ] | C'', in>
   \nonumber\\
   && - \cdots
   \nonumber\\
   &&
  - < B'', out | T [ {\bar{\eta}}^{a(m)} (x_1) F^{b(l)} (x_2) \cdots
   \delta_B F^{\hat{a}(n-{1\over2})} (x_j) \cdots ] | C'', in>
   \nonumber\\
   && - \cdots
\end{eqnarray}
leads to the Slavnov-Taylor identity
\begin{eqnarray}
\label{62}
0 &=&
  < B'', out | T [ F^{a(m)} (x_1) F^{b(l)} (x_2) \cdots
   F^{\hat{a}(n-{1\over2})} (x_j) \cdots ] | C'', in>
   \nonumber\\
   &&
  - i \delta^{ab} \delta^{ml} \delta(x_1 - x_2) < B'', out | T [ \cdots
   F^{\hat{a}(n-{1\over2})} (x_j) \cdots ] | C'', in>
   \nonumber\\
   && - \cdots
\end{eqnarray}
{\it i.e.}
\begin{eqnarray}
\label{63}
< B'', out | T [ F^{a(m)} (x_1) \cdots F^{\hat{a}(n-{1\over2})} (x_j) 
\cdots ] | C'',
in>_{con} = 0.
\end{eqnarray}
Then we expect at sufficiently high energies
\begin{eqnarray}
\label{64}
(-i) \cdots (-i) S [C'' {\to} B''
   + A_{\mu}^{a(m)} (p, {\lambda} = L) + \cdots
   + A_{\mu}^{\hat{a}(n-{1\over2})} (p', {\lambda} = L) + \cdots ]
   \nonumber\\
   = S [C'' {\to} B''
   + A_{5}^{a(m)} (p) + \cdots
   + A_{5}^{\hat{a}(n-{1\over2})} (p') + \cdots ]
   \nonumber\\
   + O \left[  \frac{M_{m}}{E}\;  \right] + \cdots +
   O \left[  \frac{M_{n-{1\over2}}}{E'}\;  \right] + \cdots ,
\end{eqnarray}
   which exibit the equivalence theorem.
   It should be noted that the 5th gauge field is
   the would-be NG-like field
   in the extra dimensional gauge theory
   where the gauge symmetry is broken
   by the nontrivial boundary conditions.

%%%%%%%%%%%%%%%%%%%%%%%%%%%%%%%%%%%%%%%%%%%%%%%%%%%%%%%%%%%%%%%%%%%
%%%%%%%%%%%%%%%%%%%%% SECTION %%%%%%%%%%%%%%%%%%%%%%%%%%%%%%%%%%%%%
%%%%%%%%%%%%%%%%%%%%%%%%%%%%%%%%%%%%%%%%%%%%%%%%%%%%%%%%%%%%%%%%%%%
\section{Amplitudes of four massive gauge bosons in the GUT on orbifold}

In the previous paper\cite{abe}
   we have calculated the amplitudes of the process
   including four massive gauge bosons in the external lines,
   in the framework of the 5-dimensional $SU(5)$ GUT,
   whose 5th-dimensional coordinate
   is compactified on $S^1/Z_2$.
We have shown that the power behavior of
    $O(E^4/m^4)$ and that of $O(E^2/m^2)$
   in the amplitude both vanish,
   and that the broken gauge theory through
   the orbifolding preserves unitarity at high energy.
Here we present the results of the calculation
   of the Feynman diagrams to $O(1)$
   and compare them with the ones for the connected reactions
   where the gauge fields are replaced
   by the corresponding would-be NG-like fields.

First, we show the notation
  of massive vector boson in the external line.
In the center-of-mass frame,
  we take the initial momentum as
\begin{eqnarray}
\label{65}
p_1=E(1,0,0,\sqrt{1-{m^2\over E^2}})
\end{eqnarray}
  and
\begin{eqnarray}
\label{66}
p_2=E(1,0,0,- \sqrt{1-{m^2\over E^2}}) ,
\end{eqnarray}
  and the final momentum as
\begin{eqnarray}
\label{67}
k_1=E(1,\sqrt{1-{m^2\over E^2}}\sin\theta,0,
           \sqrt{1-{m^2\over E^2}}\cos\theta)
\end{eqnarray}
  and
\begin{eqnarray}
\label{68}
k_2=E(1,- \sqrt{1-{m^2\over E^2}}\sin\theta,0,
           - \sqrt{1-{m^2\over E^2}}\cos\theta) ,
\end{eqnarray}
  where $m$ is the gauge boson mass.
Then the longitudinal polarization vectors
  become
\begin{eqnarray}
\label{69}
\epsilon_L(p_1)={E \over m}(\sqrt{1-{m^2\over E^2}},0,0,1) ,
\end{eqnarray}
\begin{eqnarray}
\label{70}
\epsilon_L(p_2)={E \over m}(\sqrt{1-{m^2\over E^2}},0,0,-1) ,
\end{eqnarray}
\begin{eqnarray}
\label{71}
\epsilon_L(k_1)={E \over m}(\sqrt{1-{m^2\over E^2}},
     \sin\theta,0,\cos\theta)
\end{eqnarray}
  and
\begin{eqnarray}
\label{72}
\epsilon_L(k_2)={E \over m}(\sqrt{1-{m^2\over E^2}},
     - \sin\theta,0,- \cos\theta) .
\end{eqnarray}

Before examining
  the amplitudes of four massive gauge bosons in the orbifold model,
  let us show briefly those in the 4D SM,
  where the $W$ and $Z$ gauge bosons obtain
  their masses through the Higgs mechanism.
For the process of
  $W^+ W^- \rightarrow W^+ W^-$,
  there are five diagrams:
  (1a) s-channel photon and $Z$ exchange,
  (1b) t-channel photon and $Z$ exchange,
  (1c) quadrilinear vertex,
  (1d) s-channel Higgs exchange and
  (1e) t-channel Higgs exchange.
As shown in the previous paper\cite{abe}
   the power behavior of $O(E^4/m^4)$ in the amplitude vanish
   due to the cancellation among three diagrams (1a), (1b) and (1c).
Furthermore, the power behavior of $O(E^2/m^2)$ vanish
   due to the cancellation among all diagrams
   (1a), (1b), (1c), (1d) and (1e).
In Table.1, we summarize the previous results,
   adding the new results on the amplitudes of $O(1)$.
%
%
%
%\vskip .5cm
\begin{table}[h]
\caption[table-1]{The coefficients of the
amplitude of $W^+W^-\rightarrow W^+W^-$ in
  (1a)--(1e) in the 4D SM.
Both $O(E^4/m^4)$ and $O(E^2/m^2)$ contributions
  are canceled among (1a)--(1e).}
\begin{center}
\begin{tabular}{|c||c|c|c|}     \hline
\multicolumn{1}{|c|} {}& {${ig^2 {E^4 \over m^4} \times}$}
         & {${ig^2 {E^2 \over m^2} \times}$}
         & {${ig^2 \times}$}  \\ \hline
\multicolumn{1}{|c|} {(1a)} & {$-4\cos \theta$} &
{$-\cos \theta$} &
{$3\cos \theta - {1 \over 4 \cos^2 \theta_W}\cos \theta $} \\ \hline
\multicolumn{1}{|c|} {(1b)} & {$3-2\cos \theta -\cos^2\theta$} &
{$- {3 \over 2} + {15 \over 2} \cos \theta$} &
{$- {1 \over 2} - {5 \over 2} \cos \theta + {1 \over 4 \cos^2 \theta_W} 
{3+\cos \theta \over 1-\cos \theta} $} \\ \hline
\multicolumn{1}{|c|} {(1c)} & {$-3+6\cos \theta +\cos^2\theta$} &
{$2-6\cos \theta$} &
{$-$} \\ \hline
\multicolumn{1}{|c|} {(1d)} & {$-$} &
{$-1$} &
{$1 - {1 \over 4} {m_H^2 \over m^2} $} \\ \hline
\multicolumn{1}{|c|} {(1e)} & {$-$} &
{${1 \over 2} - {1 \over 2} \cos \theta$} &
{$-{1 \over 2} - {1 \over 2} \cos \theta - {1 \over 4} {m_H^2 \over m^2} $} 
\\ \hline
\end{tabular}
\end{center}
\label{table-1}
\end{table}

On the other hand, the amplitudes for the connected reactions
   where the gauge fields $W$ are replaced
   by the corresponding would-be NG fields $G$
   are obtained as follows.
For the process of
  $G^+ G^- \rightarrow G^+ G^-$,
  there are three diagrams of $O(1)$:
  (2a) s-channel photon and $Z$ exchange,
  (2b) t-channel photon and $Z$ exchange,
  (2c) quadrilinear vertex.
In Table.2, we give the results on the amplitudes of $O(1)$.
As expected from the equivalence theorem,
   the scattering amplitude of the gauge fields
   $W^+W^-\rightarrow W^+W^-$
   coincides with the one of
   the corresponding would-be NG fields
   $G^+G^-\rightarrow G^+G^-$
   up to $O(m/E)$.
%
%
%
%\vskip .5cm
\begin{table}[h]
\caption[table-2]{The coefficients of the
amplitude of $G^+G^-\rightarrow G^+G^-$ in
  (2a)--(2c) in the 4D SM.}
\begin{center}
\begin{tabular}{|c||c|c|c|}     \hline
\multicolumn{1}{|c|} {}& {${ig^2 {E^4 \over m^4} \times}$}
         & {${ig^2 {E^2 \over m^2} \times}$}
         & {${ig^2 \times}$}  \\ \hline
\multicolumn{1}{|c|} {(2a)}  &{$-$} &{$-$} & {$ - {1 \over 4 \cos^2 
\theta_W}\cos \theta $} \\ \hline
\multicolumn{1}{|c|} {(2b)} &{$-$} &{$-$} & {$ {1 \over 4 \cos^2 \theta_W} 
{3+\cos \theta \over 1-\cos \theta} $} \\ \hline
\multicolumn{1}{|c|} {(2c)} &{$-$} &{$-$} & {$ - {1 \over 2} {m_H^2 \over 
m^2} $} \\ \hline
\end{tabular}
\end{center}
\label{table-2}
\end{table}

Next,
  let us consider
  the amplitudes of four massive gauge bosons in the 4D $SU(5)$ GUT,
  where the $X$ and $Y$ gauge bosons obtain
  masses through the Higgs mechanism.
For the process of
  $X X^{*} \rightarrow X X^{*}$,
  there are five diagrams:
  (3a) s-channel $A_3$, $A_8$, $W_3$, $B$ exchange,
  (3b) t-channel $A_3$, $A_8$, $W_3$, $B$ exchange,
  (3c) quadrilinear vertex,
  (3d) s-channel $\Sigma_3$, $\Sigma_8$, $\Sigma_{W_3}$, $\Sigma_B$ exchange and
  (3e) t-channel $\Sigma_3$, $\Sigma_8$, $\Sigma_{W_3}$, $\Sigma_B$ exchange.
Here, $A_3$, $A_8$, $W_3$ and $B$
  stand for the diagonal
  elements of the gauge fields of the $SU(5)$,
  corresponding to $SU(3)_c$, $SU(3)_c$, $SU(2)_L$,
  and $U(1)_Y$ components, respectively.
Similarly, $\Sigma_3$, $\Sigma_8$, $\Sigma_{W_3}$ and $\Sigma_B$
  stand for the diagonal
  elements of the adjoint Higgs fields of the $SU(5)$.
As shown in the previous paper\cite{abe}
   the power behavior of $O(E^4/m^4)$ in the amplitude vanish
   due to the cancellation among three diagrams (3a), (3b) and (3c).
Furthermore, the power behavior of $O(E^2/m^2)$ vanish
   due to the cancellation among all diagrams
   (3a), (3b), (3c), (3d) and (3e).
In Table.3, we summarize the previous results,
   adding the new results on the amplitudes of $O(1)$.
There appears an ``averaged mass'' of the Higgs defined as
   $m_\Sigma^2 = {1 \over 4} (m_{\Sigma_3}^2 + {1 \over 3} m_{\Sigma_8}^2 + 
m_{\Sigma_{W_3}}^2 + {5 \over 3} m_{\Sigma_B}^2)$.
%
%
%
%\vskip .5cm
\begin{table}[h]
\caption[table-3]{The coefficients of the
amplitude of $X X^{*} \rightarrow X X^{*}$ in
  (3a)--(3e) in the 4D $SU(5)$ GUT.
Both $O(E^4/m^4)$ and $O(E^2/m^2)$ contributions
  are canceled among (3a)--(3e).}
\begin{center}
\begin{tabular}{|c||c|c|c|}     \hline
\multicolumn{1}{|c|} {}& {${ig^2 {E^4 \over m^4} \times}$}
         & {${ig^2 {E^2 \over m^2} \times}$}
         & {${ig^2 \times}$}  \\ \hline
\multicolumn{1}{|c|} {(3a)} & {$-4\cos \theta$} &
{$-$} &
{$3\cos \theta$} \\ \hline
\multicolumn{1}{|c|} {(3b)} & {$3-2\cos \theta -\cos^2\theta$} &
{$8\cos \theta$} &
{$ {1 + \cos \theta + 2 \cos^2 \theta \over 1-\cos \theta} $} \\ \hline
\multicolumn{1}{|c|} {(3c)} & {$-3+6\cos \theta +\cos^2\theta$} &
{$2-6\cos \theta$} &
{$-$} \\ \hline
\multicolumn{1}{|c|} {(3d)} & {$-$} &
{$-4$} &
{$4 - {m_\Sigma^2 \over m^2} $} \\ \hline
\multicolumn{1}{|c|} {(3e)} & {$-$} &
{$2 - 2 \cos \theta$} &
{$-2 - 2 \cos \theta - {m_\Sigma^2 \over m^2} $} \\ \hline
\end{tabular}
\end{center}
\label{table-3}
\end{table}

On the other hand, the amplitudes for the connected reactions
   where the gauge fields $X$ are replaced
   by the corresponding would-be NG fields $G$
   are obtained as follows.
For the process of
  $G G^{*} \rightarrow G G^{*}$,
  there are three diagrams of $O(1)$:
  (4a) s-channel $A_3$, $A_8$, $W_3$, $B$ exchange,
  (4b) t-channel $A_3$, $A_8$, $W_3$, $B$ exchange,
  (4c) quadrilinear vertex.
In Table.4, we give the results on the amplitudes of $O(1)$.
It is noted that the coefficient of the amplitude (4c)
   is the quadrilinear coupling 
   $- 2ig^2 {m_\Sigma^2 \over m^2} = -i(\lambda_1 + {1 \over 2} \lambda_2)$.
As expected from the equivalence theorem,
   the scattering amplitude of the gauge fields
   $X X^{*} \rightarrow X X^{*}$
   coincides with the one of
   the corresponding would-be NG fields
   $G G^{*} \rightarrow G G^{*}$
   up to $O(m/E)$.
%
%
%
%\vskip .5cm
\begin{table}[h]
\caption[table-4]{The coefficients of the
amplitude of $G G^{*} \rightarrow G G^{*}$ in
  (4a)--(4c) in the 4D $SU(5)$ GUT.}
\begin{center}
\begin{tabular}{|c||c|c|c|}     \hline
\multicolumn{1}{|c|} {}& {${ig^2 {E^4 \over m^4} \times}$}
         & {${ig^2 {E^2 \over m^2} \times}$}
         & {${ig^2 \times}$}  \\ \hline
\multicolumn{1}{|c|} {(4a)}  &{$-$} &{$-$} & {$ - \cos \theta $} \\ \hline
\multicolumn{1}{|c|} {(4b)} &{$-$} &{$-$} & {$ {3+\cos \theta \over 1-\cos 
\theta} $} \\ \hline
\multicolumn{1}{|c|} {(4c)} &{$-$} &{$-$} & {$ - 2 {m_\Sigma^2 \over m^2} 
$} \\ \hline
\end{tabular}
\end{center}
\label{table-4}
\end{table}

Finaly,
  we examine the 5D $SU(5)$ theory with
  the $Z_2$ parity operator, $P'= diag.(-1,-1,-1,1,1)$,
  which realizes the gauge reduction of
  $SU(5) \rightarrow SU(3)_c \times SU(2)_L\times U(1)_Y$ \cite{5d}.
For the process of
  $X^{(1/2)} X^{(1/2)*} \rightarrow X^{(1/2)} X^{(1/2)*}$,
  there are five diagrams:
  (5a) s-channel $A_3^{(0)}$, $A_8^{(0)}$, $W_3^{(0)}$, $B^{(0)}$ exchange,
  (5b) s-channel $A_3^{(1)}$, $A_8^{(1)}$, $W_3^{(1)}$, $B^{(1)}$ exchange,
  (5c) t-channel $A_3^{(0)}$, $A_8^{(0)}$, $W_3^{(0)}$, $B^{(0)}$ exchange,
  (5d) t-channel $A_3^{(1)}$, $A_8^{(1)}$, $W_3^{(1)}$, $B^{(1)}$ exchange and
  (5e) quadrilinear vertex.
Here, $A_3^{(0)}$, $A_8^{(0)}$, $W_3^{(0)}$, $B^{(0)}$
  and $A_3^{(1)}$, $A_8^{(1)}$, $W_3^{(1)}$, $B^{(1)}$
  stand for the zero modes and the KK excited modes, respectively.
As shown in the previous paper\cite{abe}
   the power behavior of $O(E^4/m^4)$ and $O(E^2/m^2)$ vanishes
   due to the cancellation among diagrams
   (5a), (5b), (5c), (5d) and (5e).
In Table.5, we summarize the previous results,
   adding the new results on the amplitudes of $O(1)$.
%
%
%
%\vskip .5cm
\begin{table}[h]
\caption[table-5]{The coefficients of the
amplitude of $X^{(1/2)} X^{(1/2)*} \rightarrow X^{(1/2)} X^{(1/2)*}$ in
  (5a)--(5e) in the 5D $SU(5)$ GUT.
Both $O(E^4/m^4)$ and $O(E^2/m^2)$ contributions
  are canceled among (5a)--(5e).}
\begin{center}
\begin{tabular}{|c||c|c|c|}     \hline
\multicolumn{1}{|c|} {}& {${ig^2 {E^4 \over m^4} \times}$}
         & {${ig^2 {E^2 \over m^2} \times}$}
         & {${ig^2 \times}$}  \\ \hline
\multicolumn{1}{|c|} {(5a)} & {$-4\cos \theta$} &
{$-$} &
{$3\cos \theta$} \\ \hline
\multicolumn{1}{|c|} {(5b)} & {$-2\cos \theta$} &
{$-2\cos \theta$} &
{$-{1 \over 2}\cos \theta$} \\ \hline
\multicolumn{1}{|c|} {(5c)} & {$3-2\cos \theta -\cos^2\theta$} &
{$8\cos \theta$} &
{$ {1 + \cos \theta + 2 \cos^2 \theta \over 1-\cos \theta} $} \\ \hline
\multicolumn{1}{|c|} {(5d)} & {${3 \over 2}-\cos \theta -{1 \over 
2}\cos^2\theta$} &
{$-3+3\cos \theta$} &
{$ {1 \over 2}{7 - 7 \cos \theta + 4 \cos^2 \theta \over 1-\cos \theta} $} 
\\ \hline
\multicolumn{1}{|c|} {(5e)} & {$-{9 \over 2}+9\cos \theta +{3 \over 
2}\cos^2\theta$} &
{$3-9\cos \theta$} &
{$-$} \\ \hline
\end{tabular}
\end{center}
\label{table-5}
\end{table}

In the above calculations, we remind that
  the 5D gauge coupling, $g_5$,
  has mass dimension $-1/2$,
  which relates the 4D gauge coupling, $g_4=g$,
  as $g_4=g_5/\sqrt{2\pi R}$.
Then
  the couplings of $A_3^{(0)}-X^{(1/2)}-X^{(1/2)*}$,
  $A_8^{(0)}-X^{(1/2)}-X^{(1/2)*}$,
  $W_3^{(0)}-X^{(1/2)}-X^{(1/2)*}$ and
  $B^{(0)}-X^{(1/2)}-X^{(1/2)*}$ are
\begin{eqnarray}
\label{g73}
g_5 \int_{0}^{2\pi R} dy
  \left({1\over \sqrt{\pi R}}\cos{y \over 2R} \right)^2 {1\over
  \sqrt{2\pi R}}=g_4,
\end{eqnarray}
while
  the couplings of $A_3^{(1)}-X^{(1/2)}-X^{(1/2)*}$,
  $A_8^{(1)}-X^{(1/2)}-X^{(1/2)*}$,
  $W_3^{(1)}-X^{(1/2)}-X^{(1/2)*}$ and
  $B^{(1)}-X^{(1/2)}-X^{(1/2)*}$ are
\begin{eqnarray}
\label{g74}
g_5 \int_{0}^{2\pi R} dy
  \left({1\over \sqrt{\pi R}}\cos{y \over 2R} \right)^2
  \left({1\over \sqrt{\pi R}}\cos{y \over R} \right)
  ={g_4 \over \sqrt{2}}.
\end{eqnarray}
As for the four point vertex in the process (5e),
  the coupling is
  given by
\begin{eqnarray}
\label{g75}
i g_5^2 \int_{0}^{2\pi R} dy
  \left({1\over \sqrt{\pi R}}\cos{y \over 2R} \right)^4
  ={3\over2}i g_4^2.
\end{eqnarray}
This means that
  the amplitude of (5e) becomes
  3/2 times as large as that of (3c).
The  power behavior of $O(E^4/m^4)$ and that of $O(E^2/m^2)$ are both canceled
  although there are no Higgs contributed diagrams
  as the 4D $SU(5)$ GUT ((3d),(3e)).
Table 5 suggests that
  the KK-modes play the important
  rolls for preserving the unitarity,
  as noted in the previous paper\cite{abe}.
They realize the cancellation of
  the power behavior of $O(E^2/m^2)$
  as the Higgs scalars do
  in the spontaneous
  breaking gauge theories as well as
  that of $O(E^4/m^4)$.

On the other hand, the amplitudes for the connected reactions
   where the 4D gauge fields $X^{(1/2)}$ are replaced
   by the 5th gauge fields $X_5^{(1/2)}$
   are obtained as follows.
For the process of
  $X_5^{(1/2)} X_5^{(1/2)*} \rightarrow X_5^{(1/2)} X_5^{(1/2)*}$,
  there are four diagrams of $O(1)$:
  (6a) s-channel $A_3^{(0)}$, $A_8^{(0)}$, $W_3^{(0)}$, $B^{(0)}$ exchange,
  (6b) s-channel $A_3^{(1)}$, $A_8^{(1)}$, $W_3^{(1)}$, $B^{(1)}$ exchange,
  (6c) t-channel $A_3^{(0)}$, $A_8^{(0)}$, $W_3^{(0)}$, $B^{(0)}$ exchange and
  (6d) t-channel $A_3^{(1)}$, $A_8^{(1)}$, $W_3^{(1)}$, $B^{(1)}$ exchange.
It is noted that
  the couplings of $A_3^{(0)}-X_5^{(1/2)}-X_5^{(1/2)*}$,
  $A_8^{(0)}-X_5^{(1/2)}-X_5^{(1/2)*}$,
  $W_3^{(0)}-X_5^{(1/2)}-X_5^{(1/2)*}$ and
  $B^{(0)}-X_5^{(1/2)}-X_5^{(1/2)*}$ are
\begin{eqnarray}
\label{g76}
g_5 \int_{0}^{2\pi R} dy
  \left({1\over \sqrt{\pi R}}\sin{y \over 2R} \right)^2 {1\over
  \sqrt{2\pi R}}=g_4,
\end{eqnarray}
while
  the couplings of $A_3^{(1)}-X_5^{(1/2)}-X_5^{(1/2)*}$,
  $A_8^{(1)}-X_5^{(1/2)}-X_5^{(1/2)*}$,
  $W_3^{(1)}-X_5^{(1/2)}-X_5^{(1/2)*}$ and
  $B^{(1)}-X_5^{(1/2)}-X_5^{(1/2)*}$ are
\begin{eqnarray}
\label{g77}
g_5 \int_{0}^{2\pi R} dy
  \left({1\over \sqrt{\pi R}}\sin{y \over 2R} \right)^2
  \left({1\over \sqrt{\pi R}}\cos{y \over R} \right)
  =-{g_4 \over \sqrt{2}}.
\end{eqnarray}
In Table.6, we give the results on the amplitudes of $O(1)$.
As expected from the 4D equivalence theorem
   mentioned in the previous sections,
   the scattering amplitude of the 4D gauge fields
   $X^{(1/2)} X^{(1/2)*} \rightarrow X^{(1/2)} X^{(1/2)*}$
   coincides with the one of
   the 5th gauge fields {\it i.e.}
   the corresponding would-be NG-like fields
   $X_5^{(1/2)} X_5^{(1/2)*} \rightarrow X_5^{(1/2)} X_5^{(1/2)*}$
   up to $O(m/E)$ corrections.
%
%
%
%\vskip .5cm
\begin{table}[h]
\caption[table-6]{The coefficients of the
amplitude of $X_5^{(1/2)} X_5^{(1/2)*} \rightarrow X_5^{(1/2)} X_5^{(1/2)*}$ in
  (6a)--(6d) in the 5D $SU(5)$ GUT.}
\begin{center}
\begin{tabular}{|c||c|c|c|}     \hline
\multicolumn{1}{|c|} {}& {${ig^2 {E^4 \over m^4} \times}$}
         & {${ig^2 {E^2 \over m^2} \times}$}
         & {${ig^2 \times}$}  \\ \hline
\multicolumn{1}{|c|} {(6a),(6b)}  &{$-$} &{$-$} & {$ - {3 \over 2} \cos 
\theta $} \\ \hline
\multicolumn{1}{|c|} {(6c),(6d)} &{$-$} &{$-$} & {$ {3 \over 2} {3+\cos 
\theta \over 1-\cos \theta} $} \\ \hline
\end{tabular}
\end{center}
\label{table-6}
\end{table}
%
%
%

%%%%%%%%%%%%%%%%%%%%%%%%%%%%%%%%%%%%%%%%%%%%%%%%%%%%%%%%%%%%%%%%%%%
%%%%%%%%%%%%%%%%%%%%% SECTION %%%%%%%%%%%%%%%%%%%%%%%%%%%%%%%%%%%%%
%%%%%%%%%%%%%%%%%%%%%%%%%%%%%%%%%%%%%%%%%%%%%%%%%%%%%%%%%%%%%%%%%%%
\section{Summary and discussion}

In this paper we have investigated
  the high-energy behavior
  of the tree-level scattering amplitudes
  of the massive gauge bosons
  in the 5D orbifold model
  compactified on  $S^1/Z_2$.

Feynman diagram with massive vector bosons
  in external lines, in general,
  gives $O(E^4/m^4)$ and/or $O(E^2/m^2)$
  contribution to the amplitude
  and this energy-dependence could cause
  the violation of the unitarity bound.
If the mass of the gauge boson comes
  from the spontaneous breaking,
  it is well-known that this unitarity-violating
  $O(E^4/m^4)$ and $O(E^2/m^2)$ contributions
  are canceled among diagrams
  and furthermore the amplitude of $O(1)$
  is the same as the scattering amplitude
  of the corresponding would-be NG-like fields
  up to some constant factor.
This equivalence between the amplitudes comes from
  the gauge invariance of the theory.
The orbifold model, on the other hand,
  the symmetry breaking occurs
  through nontrivial boundary condition
  and the boundary condition itself
  does not respect the symmetry.
In this sense it can be said
  that the orbifold model violates
  the symmetry {\it by-hand}.
Thus it is a nontrivial problem
  whether unitarity bound is maintained
  and the equivalence theorem holds or not.
We have discussed this issue in this paper.

We have first noticed in section 2
  that the 4D theory written
  in terms of the KK gauge and scalar fields
  have an invariance under the 4D gauge transformation
  which mixes the infinitely-many KK modes.
Then we have in section 3
  carried this 4D gauge symmetry to BRST formalism
  and in section 4
  derived the Slavnov-Taylor identities among amplitudes,
  which in turn gives the equivalence theorem.
Furthermore in section 5
  we have calculated the $O(1)$ amplitude
  in the 5D $SU(5)$ orbifold model
  as well as in the 4D $SU(5)$ (ordinary) GUT model.
The result confirms explicitly the equivalence theorem.

Some comments related to the unitarity are in order.

Unitarity is closely connected to renormalizability:
Renomalizable theory seems to preserve unitarity ---
  at least known renomalizable theories preserve unitarity.
However, the 4D gauge theory we have discussed
  is not a renormalizable theory
  because it consists of infinitely-many KK fields
  and whose contribution to the loop diagrams yields divergence.
The sum of the KK fields in the 4D theory
  corresponds to momentum integration
  along the extra-dimensional direction in 5D theory.
Non-renomalizabilty of our 4D theory
  originates from that of the 5D theory.
Then, what are the implications of the tree-level unitarity
  we have just shown to hold in this paper?
There are two alternatives senarios:
First, the orbifold theory is
  a low energy effective theory
  applicable to the energy less than some scale M
  and the infinitely high KK tower
  should be truncated at this scale M.
The loop diagrams cease to diverge
  owing to this KK tower truncation,
  which corresponds in 5D theory to the momentum cut-off M
  along the extra-dimensional direction.
Second scenario is to modify the orbifold theory
  to being applicable to all energy.
If we consider some physical object
  such as D-branes sitting on the orbifold fixed-point,
  the quantum fluctuation of this object
  cause the modification of our orbifold model.
The fluctuation is expected
  to yields damping factor to higher KK mode
  and make the sum over all KK mode converge\cite{Bando}.
This means that the theory has a chance to become renomalizable.
This line of thought is worth further research.

%%%%% ACKNOWLEDGEMENTS %%%%%%%%%%%%%%%%%%%%%%%%%%%%%%%%%%%

\section*{Acknowledgements}

N. H. is supported by Scientific Grants from
  the Ministry of Education and Science of Japan,
  Grant No. 14039207,
  Grant No. 14046208,  Grant No. 14740164.

%%%%% REFERENCES %%%%%%%%%%%%%%%%%%%%%%%%%%%%%%%%%%%%%%%%%

\vskip 1.5cm

% A useful Journal macro
\def\jnl#1#2#3#4{{#1}{\bf #2} (#4) #3}

\def\Zphys{{\em Z.\ Phys.} }
\def\jssc{{\em J.\ Solid State Chem.\ }}
\def\jpsJ{{\em J.\ Phys.\ Soc.\ Japan }}
\def\ptps{{\em Prog.\ Theoret.\ Phys.\ Suppl.\ }}
\def\PTP{{\em Prog.\ Theoret.\ Phys.\  }}

\def\JMP{{\em J. Math.\ Phys.} }
\def\NPB{{\em Nucl.\ Phys.} B}
\def\NP{{\em Nucl.\ Phys.} }
\def\PLB{{\em Phys.\ Lett.} B}
\def\PL{{\em Phys.\ Lett.} }
\def\PRL{\em Phys.\ Rev.\ Lett. }
\def\PRB{{\em Phys.\ Rev.} B}
\def\PRD{{\em Phys.\ Rev.} D}
\def\PRe{{\em Phys.\ Rep.} }
\def\AP{{\em Ann.\ Phys.\ (N.Y.)} }
\def\RMP{{\
em Rev.\ Mod.\ Phys.} }
\def\ZPC{{\em Z.\ Phys.} C}
\def\SCI{\em Science}
\def\CMP{\em Comm.\ Math.\ Phys. }
\def\MPLA{{\em Mod.\ Phys.\ Lett.} A}
\def\IJMPA{{\em Int.\ J.\ Mod.\ Phys.} A}
\def\IJMPB{{\em Int.\ J.\ Mod.\ Phys.} B}
\def\EPJC{{\em Eur.\ Phys.\ J.} C}
\def\PR{{\em Phys.\ Rev.} }
\def\JHEP{{\em JHEP} }
\def\cmp{{\em Com.\ Math.\ Phys.}}
\def\JPA{{\em J.\  Phys.} A}
\def\CQG{\em Class.\ Quant.\ Grav. }
\def\ATMP{{\em Adv.\ Theoret.\ Math.\ Phys.} }
\def\ibid{{\em ibid.} }

\leftline{\bf References}

\renewenvironment{thebibliography}[1]
           {\begin{list}{[$\,$\arabic{enumi}$\,$]}  % {\arabic{enumi}.}
           {\usecounter{enumi}\setlength{\parsep}{0pt}
            \setlength{\itemsep}{0pt}  \renewcommand{\baselinestretch}{1.2}
            \settowidth
           {\labelwidth}{#1 ~ ~}\sloppy}}{\end{list}}


\begin{thebibliography}{99}
\small
\baselineskip=14pt




%%%%%%%%%%%%%%%%%%%%%%%%%%%%%%%%%%%%%%%%%%%%%%%%%%%%%%%%%%%%%
% Some macros are available for the bibliography:
%  o for general use
%    \JL : general journals                 \andvol : Vol (Year) Page
%  o for individual journal
%    \AJ   : Astrophys. J.           \NC         : Nuovo Cim.
%    \ANN  : Ann. of Phys.           \NPA, \NPB  : Nucl. Phys. [A,B]
%    \CMP  : Commun. Math. Phys.     \PLA, \PLB  : Phys. Lett. [A,B]
%    \IJMP : Int. J. Mod. Phys.      \PRA - \PRE : Phys. Rev. [A-E]
%    \JHEP : J. High Energy Phys.    \PRL        : Phys. Rev. Lett.
%    \JMP  : J. Math. Phys.          \PRP        : Phys. Rep.
%    \JP   : J. of Phys.             \PTP        : Prog. Theor. Phys.
%    \JPSJ : J. Phys. Soc. Jpn.      \PTPS       : Prog. Theor. Phys. Suppl.
% Usage:
%  \PR{D45,1990,345}            ==> Phys.~Rev.\ \textbf{D45} (1990), 345
%  \JL{Phys.~Lett.,A30,1981,56} ==> Phys.~Lett.\ \textbf{A30} (1981), 56
%  \andvol{B123,1995,1020}      ==> \textbf{B123} (1995), 1020
%%%%%%%%%%%%%%%%%%%%%%%%%%%%%%%%%%%%%%%%%%%%%%%%%%%%%%%%%%%%%






%%%%%%%%%%%%%%%%%%%%%%%%%%%%%%%%%%%%%%%%%%%%%%
%%%%%%%%%%% 7/19/2003 by NH %%%%%%%%%%%%%%%%%%
%%%%%%%%%%%%%%%%%%%%%%%%%%%%%%%%%%%%%%%%%%%%%%
\bibitem{LlewellynSmith}
C.~H.~Llewellyn Smith,
%``High-Energy Behavior And Gauge Symmetry,''
Phys.\ Lett.\ B {\bf 46}, 233 (1973).
%%CITATION = PHLTA,B46,233;%%

\bibitem{Dicus}
D.~A.~Dicus and V.~S.~Mathur,
%``Upper Bounds On The Values Of Masses In Unified Gauge Theories,''
Phys.\ Rev.\ D {\bf 7}, 3111 (1973).
%%CITATION = PHRVA,D7,3111;%%


\bibitem{GBequivalence}
J.~M.~Cornwall, D.~N.~Levin and G.~Tiktopoulos,
%``Uniqueness Of Spontaneously Broken Gauge Theories,''
Phys.\ Rev.\ Lett.\  {\bf 30}, 1268 (1973)
[Erratum-ibid.\  {\bf 31}, 572 (1973)]
%%CITATION = PRLTA,30,1268;%%
%
and
%``Derivation Of Gauge Invariance From High-Energy Unitarity Bounds On The 
S - Matrix,''
Phys.\ Rev.\ D {\bf 10}, 1145 (1974)
[Erratum-ibid.\ D {\bf 11}, 972 (1975)].
%%CITATION = PHRVA,D10,1145;%%

\bibitem{SMunitarity}
B.~W.~Lee, C.~Quigg and H.~B.~Thacker,
%``The Strength Of Weak Interactions At Very High-Energies And The Higgs 
Boson Mass,''
Phys.\ Rev.\ Lett.\  {\bf 38}, 883 (1977)
%%CITATION = PRLTA,38,883;%%
%
and
%B.~W.~Lee, C.~Quigg and H.~B.~Thacker,
%``Weak Interactions At Very High-Energies: The Role Of The Higgs Boson Mass,''
Phys.\ Rev.\ D {\bf 16}, 1519 (1977).
%%CITATION = PHRVA,D16,1519;%%

\bibitem{Chanowitz}
M.~S.~Chanowitz and M.~K.~Gaillard,
%``The Tev Physics Of Strongly Interacting W's And Z's,''
Nucl.\ Phys.\ B {\bf 261}, 379 (1985).
%%CITATION = NUPHA,B261,379;%%






%%%%%%%%%%%%%%%%%%%%%%%%%%%%%%%%%%%%%%%%%%%%%%
%%%%%%%%%%% 7/21/2003 by YA %%%%%%%%%%%%%%%%%%
%%%%%%%%%%%%%%%%%%%%%%%%%%%%%%%%%%%%%%%%%%%%%%

\bibitem{Gounaris}
G.~J.~Gounaris, R.~K$\ddot{{\rm o}}$gerler and H.~Neufeld,
%``Relationship between longitudinally polarized vector''
Phys.\ Rev.\ D {\bf 34}, 3257 (1986).

\bibitem{Yao}
Y.~-P.~Yao and C.~-P.~Yuan,
%``Modification of the equivalence theorem''
Phys.\ Rev.\ D {\bf 38}, 2237 (1988).

\bibitem{Bagger}
J.~Bagger and C.~Schmidt,
%``Equivalence theorem redux
Phys.\ Rev.\ D {\bf 41}, 264 (1990).

\bibitem{He}
H.~-J.~He, Y.~-P.~Kuang and X.~Li,
%``On the precise formulation of the equivalence theorem
Phys.\ Rev.\ Lett.\  {\bf 69}, 2619 (1992) and
%``Fuether investigation on the precise formulation''
Phys.\ Rev.\ D {\bf 49}, 4842 (1994).

%%%%%%%%%%%%%%%%%%%%%%%%%%%%%%%%%%%%%%%%%%%%%%
%%%%%%%%%%%%%%%%%%%%%%%%%%%%%%%%%%%%%%%%%%%%%%
%%%%%%%%%%%%%%%%%%%%%%%%%%%%%%%%%%%%%%%%%%%%%%



\bibitem{5d}
Y.~Kawamura,
Prog.\ Theor.\ Phys.\  {\bf 103} (2000),  613; ibid
\  {\bf 105} (2001), 691; ibid {\bf 105} (2001), 999;\\
%``Split multiplets, coupling unification and extra dimension,''
%``Triplet-doublet splitting, proton stability and extra dimension,''
G.~Altarelli and F.~Feruglio,
%``SU(5) grand unification in extra dimensions and proton decay,''
Phys.\ Lett.\ B {\bf 511} (2001), 257;\\
L.~J.~Hall and Y.~Nomura,
%``Gauge Unification in Higher Dimensions''
Phys.\ Rev. D {\bf 64} (2001), 055003.
L.~J.~Hall and Y.~Nomura,
%Gauge Coupling Unification from Unified Theories in Higher Dimensions
Phys. Rev. {\bf D65} (2002), 125012.



\bibitem{KK}
Th.~Kaluza, Sitzungsber, Preuss. Akad. Wiss. Berlin, Phys. Math. Klasse
(1921) 966;\\
O.~Klein, {\it Z. Phys.} {\bf 37} (1926) 895.




\bibitem{Sekhar1}
R.~S.~Chivukula, D.~A.~Dicus and H.~J.~He,
%``Unitarity of compactified five-dimensional Yang-Mills theory,''
Phys.\ Lett.\ B {\bf 525}, 175 (2002).
%[{\tt hep-ph/0111016}].
%%CITATION = HEP-PH 0111016;%%

\bibitem{Sekhar3}
R.~S.~Chivukula, D.~A.~Dicus, H.~J.~He and S.~Nandi,
%``Unitarity of the higher dimensional standard model,''
{hep-ph/0302263}.
%%CITATION = HEP-PH 0302263;%%

\bibitem{abe}
Y.~Abe, N.~Haba, Y.~Higashide, K.~Kobayashi, M.~Matsunaga,
Prog. Theor. Phys.{\bf 109} (2003) 831.

\bibitem{1}
M. E. Shaposhnikov and P. Tinyakov,
  Phys. Lett. {\bf B515} (2001) 442.

\bibitem{2}
R. S. Chivukula and H. -J. He,
  Phys. Lett. {\bf B532} (2002) 121.

\bibitem{3}
S.~De Curtis, D.~Dominici and J.~R.~Pelaez,
%``Strong tree level unitarity violations in the extra dimensional 
standard model with scalars in the bulk,''
Phys.\ Rev.\ D {\bf 67}, 076010 (2003).


\bibitem{uni}
A.~Hebecker and J.~March-Russell,
%``The Structure of GUT Breaking by Orbifolding''
Nucl. Phys. {\bf B613} (2002) 128.

\bibitem{DeCurtis}
S.~De Curtis, D.~Dominici and J.~R.~Pelaez,
%``The equivalence theorem for gauge boson scattering in a five 
dimensional standard model,''
Phys.\ Lett.\ B {\bf 554}, 164 (2003).
%[{\tt hep-ph/0211353}]
%%CITATION = HEP-PH 0211353;%%
and
%\bibitem{DeCurtis:2003zt}
%S.~De Curtis, D.~Dominici and J.~R.~Pelaez,
%``Strong tree level unitarity violations in the extra dimensional 
standard model with scalars in the bulk,''
Phys.\ Rev.\ D {\bf 67}, 076010 (2003).
%[{\tt hep-ph/0301059}].
%%CITATION = HEP-PH 0301059;%%

\bibitem{Hall:2001tn}
L.~J.~Hall, H.~Murayama and Y.~Nomura,
%``Wilson lines and symmetry breaking on orbifolds,''
Nucl.\ Phys.\ B {\bf 645} (2002) 85.
%[arXiv:hep-th/0107245].
%%CITATION = HEP-TH 0107245;%%




\bibitem{Schwartz}
M.~D.~Schwartz,
%``Constructing gravitational dimensions,''
{hep-th/0303114}.
%%CITATION = HEP-TH 0303114;%%



\bibitem{Murayama}
C.~Csaki, C.~Grojean, H.~Murayama, L.~Pilo and J.~Terning,
%``Gauge theories on an interval: Unitarity without a Higgs,''
hep-ph/0305237.
%%CITATION = HEP-PH 0305237;%%


\bibitem{Hall:2001zb}
L.~J.~Hall, Y.~Nomura and D.~R.~Smith,
%``Gauge-Higgs unification in higher dimensions,''
Nucl.\ Phys.\ B {\bf 639}, 307 (2002).


\bibitem{Bando}
M.~Bando, T.~Kugo, T.~Noguchi and K.~Yoshioka,
Phys.\ Rev.\ Lett. {\bf 83}, 3601 (1999).

%%%%%%%%%%%%%%%%%%%%%%%%%%%%%%%%%%%%%%%%%%%%%%
%%%%%%%%%%%%%%%%%%%%%%%%%%%%%%%%%%%%%%%%%%%%%%
%%%%%%%%%%%%%%%%%%%%%%%%%%%%%%%%%%%%%%%%%%%%%%













\end{thebibliography}
\end{document}